
\hoffset=0 true in
\hsize=6.75 true in
\voffset=0 true in
\vsize=9 true in
\overfullrule=0pt
\abovedisplayskip=15pt
\belowdisplayskip=15pt
\abovedisplayshortskip=10pt
\belowdisplayshortskip=10pt

\def\gapp{\mathrel{\raise.3ex\hbox{$>$}\mkern-14mu
              \lower0.6ex\hbox{$\sim$}}}
\def\hbar{{\mathchar'26\kern-.5em{\it h}}}



\def\gtorder{\mathrel{\raise.3ex\hbox{$>$}\mkern-14mu
             \lower0.6ex\hbox{$\sim$}}}
\def\ltorder{\mathrel{\raise.3ex\hbox{$<$}\mkern-14mu
             \lower0.6ex\hbox{$\sim$}}}

\baselineskip=14truept plus 1pt minus 2pt
$  $
\vskip 0.5truein
\centerline {\bf Flare Production of $^6$Li in Population~II Stars}
\medskip
\noindent
\centerline{Constantine P. Deliyannis$^1$ and  Robert. A. Malaney$^2$}

\medskip
\noindent
$^1$ Hubble Postdoctoral Fellow, Institute for
Astronomy, University of Hawaii, 2680 Woodlawn Dr, Honolulu, HI,
U.S.A.
96822. Email: con@galileo.ifa.hawaii.edu, con@astro.yale.edu
\medskip
\noindent
$^2$ Canadian Institute for Theoretical Astrophysics, University of
Toronto,
Toronto, ON, CANADA M5S 1A7. Email: malaney@cita.utoronto.ca

\vskip 0.5truein
\baselineskip=14truept plus 1pt minus 2pt

\centerline{ABSTRACT}

Motivated by the recent  report of a  $^6$Li detection in the atmosphere of
HD~84937,
we couple stellar evolution calculations with light isotope production
via stellar flares. We find that
as a consequence of their small convective envelope mass near the
turn-off point, low-metallicity dwarfs and subgiants may possess
observable amounts of $^6$Li arising from such flare activity.
We point out an observational test which could discriminate between
flare produced $^6$Li and protostellar $^6$Li. In the
$T_{\rm eff}$ range $6000-6600$~K, the $^7$Li/$^6$Li
ratio on the subgiant branch should increase as a function of $T_{\rm
eff}$
if flare production is important, whereas the  same ratio should be
constant
if a protostellar origin is the source of the observed lithium.
The absence of a flare produced variation in the $^7$Li/$^6$Li ratio
would allow for a more reliable inference of the -- cosmologically
important -- atmospheric depletion mechanisms in stars.

\vfill
\eject
\centerline{1. INTRODUCTION}

	The recent report of a detection of $^6$Li in the metal poor halo
star HD~84937 by Smith, Lambert, and Nissen (1993; hereafter SLN) could
potentially
be
significant for cosmology, by constraining estimates for the primordial
abundance of the heavier isotope, $^7$Li.  Knowledge of the primordial
$^7$Li abundance, ${\rm Li}_p$, constrains models of big bang
nucleosynthesis
(BBN).  In particular, ${\rm Li}_p$ plays an important role in
in testing the validity of standard BBN (eg. Krauss and Romanelli 1990;
Walker {\it et al.} 1991; Smith, Kawano and Malaney 1993).
  ${\rm Li}_p$ can
potentially be
derived from the Li abundances observed in the old halo dwarfs, which
exhibit a plateau of nearly constant Li abundance for stars with $6400 \ge
T_{\rm eff} \ge 5600$ K, and depleted Li abundances relative to the
plateau for
both cooler dwarfs and cooler subgiants.  The level of the plateau has
often been associated with ${\rm Li}_p$, consistent with standard BBN.

	However, both Li isotopes are very fragile, $^6$Li more so, and
are destroyed easily by (p,$\alpha$) reactions at only a few million
degrees.
Thus, already when stars arrive on the ZAMS, Li survives only in the
outermost few percent (by mass) of the stellar interior.  There are several
physical mechanisms that indeed could alter the surface abundances
during
the stellar lifetime.  Therefore, by necessity, a thorough understanding of
these mechanisms must precede confident
 evaluation of ${\rm Li}_p$.  At present,
at least two
classes of stellar models are able to reproduce at least some aspects of the
observations, but with different conclusions.  Standard stellar
evolutionary models (that ignore possible effects due to diffusion,
rotational mixing, mass loss, magnetic fields) reproduce the general
features of the observations with very little $^7$Li depletion in the
plateau
(Deliyannis, Demarque, and Kawaler 1990).  This implies self-consistency
in standard BBN, and possibly the necessity for non-baryonic
dark matter in galactic halos and larger
scales.  The reported detection of $^6$Li, at the level of about 1 part in 20
relative to $^7$Li, could then be consistent with its production by
Galactic
Cosmic Rays (GCR) in the interstellar medium. (GCR production of light isotopes
in the early galaxy has been widely discussed in the recent literature; eg.
Steigman and Walker 1992; Duncan, Lambert and Lemke 1992; Prantzos, Casse and
Vangionni-Flam 1993;
Malaney and Butler 1993; Feltzing and Gustafsson 1994; Fields, Olive, and
Schramm 1994).
 However, the
predictions
of standard stellar models are blatantly contradicted by Pop I data (e.g.
the
Boesgaard Li gap and depletion of Be in F stars, the degree and timing of
Li depletion in open clusters, the dispersion of Li abundances observed
in
open clusters, the higher Li abundances seen in short period tidally
locked
binaries), so the approximate agreement with the Pop II data could be
coincidental.

	On the other hand, stellar evolutionary models with rotationally
induced mixing deplete the Li plateau by at least a factor of 3 - 10
(Pinsonneault, Deliyannis, and Demarque 1992), and the resulting ${\rm
Li}_p$
introduces complications, and the possibility of some non-standard
BBN (see Malaney and Mathews 1993 for review).
  If the D$+^3$He constraint were relaxed, however (as also
possibly suggested by rotational mixing, Deliyannis {\it et al.} 1994),
consistent
with the recent possible detection of D at high redshift (Songaila {\it et
al.}
1994; Carswell {\it et al.} 1994), then standard BBN consistency could be
maintained at low
baryon
density, requiring non-baryonic dark matter in galactic halos and larger
scales.  Consistency could again be maintained if the primordial helium
abundance were just slightly higher than current estimates (and ignoring
the quasar D), in which case dark galactic halos and larger scales could
be baryonic.

	Which class of stellar models, if either, might be more realistic?
Newer models with improved internal physics have verified the
robustness of the predictions in both standard and rotational models
(Chaboyer and Demarque 1994).  Unlike standard models, the rotational
models can explain all the features of Pop I stars listed above.  As in the
case of Pop I stars, the rotational models predict a dispersion in the Li
abundances in halo stars.
A detailed analysis of the available data showed that such a dispersion
indeed exists (Deliyannis, Pinsonneault, and Duncan 1993), and with
about the right magnitude.  Thorburn (1994) has verified this dispersion
with a much larger sample of stars, though she suggested it might be
related to Galactic Li enrichment combined with an age spread in
the halo.  Debate continues while the value of ${\rm Li}_p$ remains
poorly
known.  Note that the curious linear relationship of Be/B $v$ metallicity
for the halo dwarfs (eg. Gilmore {\it et al} 1992; Boesgaard and King 1993)
might also warrant
production scenarios for the light elements ($^6$Li, $^7$Li, $^{9}$Be,
$^{10}$B, $^{11}$B) other than standard BBN$+$GCR.
 Also note that in  standard BBN, $^6$Li is at least 100 times less
abundant
than $^7$Li, and thus
currently not possible to detect reliably even if perfectly preserved in the
stellar atmosphere;  however, other models of primordial nucleosynthesis
predict much lower ratios of $^7$Li/$^6$Li (e.g. Dimopoulos {\it et
al.} 1988).

	$^6$Li could provide clues in the following manner:  Assuming the presence of
$^6$Li
in  a halo star, and if that $^6$Li was
indeed present in the material that formed the star, then the degree of
rotationally induced depletion of $^7$Li  might be usefully
constrained observationally.  However, before drawing such powerful
conclusions we should caution that  -- as pointed out by SLN
 -- contamination of the atmosphere of HD~84937 with
nucleosynthesis products arising from stellar flares is a real possibility.
The introduction of a sufficient amount of flare produced $^6$Li would
place no constraints no rotational depletion of $^7$Li.

	It is the purpose of this work, to quantitatively investigate the
hypothesis that
the $^6$Li detected in HD~84937 was produced by stellar flares.  We
also
provide predictions to distinguish between flare production and GCR production.

\medskip
\centerline{2. HD~84937; STELLAR MODELS}

	Any $^6$Li that might be produced by autogenetic surface
spallation
(hereafter ``surface spallated $^6$Li") will be diluted in the surface
convection
zone (SCZ).  Therefore, determining whether flares can produce
observable amounts of $^6$Li first requires estimating the size, in terms
of
mass, of the SCZ.  Stellar evolutionary models show that the mass, $M_c$,  of
the
SCZ depends on several model parameters: effective temperature,
metallicity, choice of mixing length; it also depends on age (Deliyannis {\it
et al.}
 1990).  Furthermore, given the fragility of $^6$Li wrt
(p,$\alpha$) reactions,
the base of the SCZ could be sufficiently hot and dense to destroy this
$^6$Li.  Indeed, detailed stellar models (Deliyannis  {\it et al.} 1989;,
1990) show
that, if halo dwarfs formed with $^6$Li (hereafter ``protostellar
$^6$Li"), this $^6$Li
could be preserved today only in stars near the turnoff.  In this section
we
use standard (e.g. no diffusion, rotation, mass loss, or other additional
physics not usually included in models) stellar evolutionary models to
examine the necessary constraints (such as minimum $M_c$ ) on
modelling of
surface spallation that must be met if flares are to have produced $^6$Li
that is
observable today in stellar atmospheres.  (Possible complications due to
dependence on parameters and additional physics is discussed in section
4.)  We relate these findings to the case of HD~84937.

	It should be kept in mind that the precise evolutionary status of
HD
84937 is not known (SLN).  For clarity, we define the
turnoff
as precisely the bluest point in a cluster;  furthermore, stars above that
point but still below the giant branch will be referred to as subgiants,
whereas stars below that point will be referred to as dwarfs.  When the
B-
V color of HD~84937 ($ \sim 0.40$; values in SIMBAD range from
0.37 to
0.42) is compared to the turnoff region of M92, which has similar
metallicity and is one of the oldest and most metal-poor globular clusters,
one finds that HD~84937 could either be a genuine dwarf just below the
turnoff (position A in Figure 1), or a subgiant just above the turnoff
(position B).  It could even be (e.g. if it is older than M92) right at the
turnoff.  Note that Demarque, Deliyannis, and Sarajedini (1991) have
derived an age of $ \sim  17$ Gyr for M92 (Figure 1; slight differences
between
the models described below and of Demarque  {\it et al.} do not affect
the
derived age); others have obtained similar ages (Chaboyer {\it et al.}
1992,
Proffitt and Vandenberg 1991, slightly higher in Straniero and Chieffi
1991).

	We begin in this section by considering the (standard) models of
Deliyannis (1990) and Deliyannis  {\it et al.} (1989, 1990), collectively
referred
to as DDK, for $Z= 10^{-4}$  (${\rm [Fe/H]}  \approx -2.3$, i.e.
similar metallicity as HD
84937), both to review survival of protostellar $^6$Li and to determine
how
much mass ($M_c$)  is to be enriched with spallated $^6$Li.
We note
that these models were able to reproduce the general morphology of the
$^7$Li observations (depleted cool dwarfs, plateau, and diluted
subgiants).

\medskip
\noindent
{\it 2.1  Brief Review of Protostellar $^6$Li Evolution}

	We will need to distinguish between protostellar $^6$Li and
surface
spallated $^6$Li, so we briefly review the possible survival of
protostellar $^6$Li
in present stellar atmospheres.  Both Li isotopes can be destroyed
($^6$Li
more easily) by (p,$\alpha$) reactions at the base of the SCZ,
 more efficiently
in deeper convection zones that are hotter at their base.  Figure 2 shows
the evolution of the mass $M_c$  contained in the SCZ for model
sequences of
different stellar mass, as a function of age (for $\alpha = 1.5$; Deliyannis
and Demarque (1991) show the corresponding curves for the temperature
and density at the base of the SCZ; see Deliyannis (1990) for a larger
parameter space).  The models begin their evolution fully convective,
high
up the Hayashi track; subsequently, the SCZ steadily becomes shallower
until the turnoff (except in the final approach to the ZAMS), and then
steadily becomes deeper during subgiant evolution (see DDK for further
details).  Most of the Li destruction occurs during the pre-main sequence,
when the convection zones are deepest, and for the lowest massed
models
(which have the deepest convection zones).  Figure 3 shows evolution of
the same model sequences in the $M_c$  - $T_{\rm eff}$ plane.  Lower
massed models
arrive on the ZAMS (diamond symbols) at lower $T_{\rm eff}$, and
preserve their
rank as they evolve on the main sequence.  For these reasons, the
amount
of protostellar $^6$Li that survives today in dwarfs decreases with
decreasing
$T_{\rm eff}$ (Figure 4a).

	Since it is difficult to distinguish observationally between field
subgiants and dwarfs near the turnoff, such as in the case of HD~84937,
we also consider subgiant evolution.  Furthermore, it will turn out that
slightly more evolved subgiants can provide a good test to distinguish
between surface spallated $^6$Li and protostellar $^6$Li.  On an
isochrone,
subgiants have evolved from turnoff $T_{\rm eff}$ just beyond (hotter
than) the
present turnoff, and as such had preserved more protostellar $^6$Li (and
$^7$Li)
than current dwarfs (that are cooler than the isochrone turnoff $T_{\rm
eff}$).  After
a model evolves past the turnoff, its SCZ begins to deepen (Figure 3).
The $^6$Li abundance remains constant until the SCZ deepens beyond
the
depth where protostellar $^6$Li was preserved on the main sequence;
subsequently, the surface $^6$Li abundance will begin to dilute (Figure
3;
for details see Deliyannis {\it et al.} 1990 and Ryan and Deliyannis 1994).
Figure 4a summarizes the predicted $^6$Li abundance surviving from a
constant protostellar $^6$Li abundance for both dwarfs and subgiants.
The
 subgiant $^7$Li/$^6$Li has been normalized
to 20  at $6250$~K, to agree with HD~84937.  If
protostellar Li has a primordial component plus roughly equal
contributions
of $^6$Li and $^7$Li produced by GCR's interacting with
the ISM, and if the observed $^7$Li abundance in HD~84937 is  $A(^7{\rm Li})  =
 12  +  {\rm log} ( n(^7{\rm Li)} / n({\rm H))}\approx  2.1  $, then this
normalization
is
possible (with the models shown) if ${\rm Li}_p  \approx  2.0$  and
$^7{\rm Li(GCR) }  \approx ^6{\rm Li(GCR)} \approx 1.5 $ .
We note that the results shown in Figure~4a depend on age,
metallicity, and other parameters (see DDK).

\medskip
\noindent
{\it 2.2.  Surface Spallated $^6$Li:  Dependence of $M_c$  on
$T_{\rm eff}$ (Mass and Age)}

	A similar figure (4b) will be presented in section 3 showing how
surface spallated $^6$Li varies as a function of $T_{\rm eff}$ for both
dwarfs and
subgiants, to be compared to Figure~4a.  Here, we evaluate how the mass
$M_c$, to be enriched with surface spallated $^6$Li in section 3, varies
with
$T_{\rm eff}$.  Consider first the dwarf models.  Figure 2 shows that at
a given
age, $M_c$  is a steep function of stellar mass: for example, at 15 Gyr,
in
going from 0.65 to $0.75 M_{\odot}$, the convection zone becomes
more than
two orders of magnitude shallower.  $M_c$  is also clearly a steep
function of
age: for example, in evolving from the ZAMS to 20 Gyr, the $0.70
M_{\odot}$
model's SCZ becomes shallower by two orders of magnitude, with
much
of this decline occurring in late stages.  In particular, from 15 to 20 Gyr
the
SCZ becomes shallower by one order of magnitude.  The dependence of
$M_c$  on age implies that, unless surface spallation rates were much
higher
in the past, any observable surface spallated $^6$Li is
more likely to have been created
recently.  Since $^6$Li is destroyed only during the pre-main
sequence, it is important to note that (recently created) spallated $^6$Li
will
indeed survive to the present.

	Fortunately, the dependence of $M_c$  on $T_{\rm eff}$ and
age are related, and
a great simplification occurs in the $T_{\rm eff}$ - $M_c$  plane (Figure
3).  After
models arrive on the ZAMS, they evolve {\it almost exactly} along the curve
defined by the ZAMS itself (curve connected by diamonds in Figure~3);
therefore, the dwarf $M_c$  vs. $T_{\rm eff}$ curve is nearly
independent of age ($M_c$
isochrones fall nearly on top of each other).  We will refer to this curve
as the ``$M_c$  evolution pathway''.  Isochrones do diverge slightly
with larger
$T_{\rm eff}$, but even at $T_{\rm eff}$ = 6400 K, 14.5 and 18 Gyr
dwarf isochrones differ
by only 0.1 dex.  This suggests that we require neither a star's mass nor
its age (both difficult to obtain) to estimate its $M_c$ , though we do
need to
know its $T_{\rm eff}$ (less difficult to obtain).  Given the likelihood
that there is
an age spread among globular clusters (Searle and Zinn 1978; Lee {\it et
al.} 1990;
Demarque {\it et al.} 1991;
Sarajedini and Demarque 1991; Vandenberg {\it et al.}
1991; Chaboyer {\it et al.} 1992) as well in field halo stars (e.g. SN3),
the near
independence of $M_c$  with age in the $T_{\rm eff}$ - $M_c$  plane is
quite fortuitous.

	$M_c$  does remain a steep function of $T_{\rm eff}$: in going
from $T_{\rm eff}$ =
6040 K to 6620 K (for $\alpha = 1.5$) the convection zone becomes two
orders of magnitude shallower.  This implies that, for spallation rates
independent of $T_{\rm eff}$, the spallated $^6$Li abundance will
increase with $T_{\rm eff}$
(section 3).  This trend is, unfortunately, similar to that expected for
survival of protostellar $^6$Li (section 2.1), and so does not allow us to
distinguish easily between spallated and protostellar $^6$Li.  However,
the
subgiants potentially can.

	Consider now subgiants.  Subgiant evolution of $M_c$  is
shown for
$0.725 M_{\odot}$ and $0.775 M_{\odot}$  models in Figures 2 and 3; these
correspond to
approximate ages of 15 and 19 Gyr.  The evolutionary timescale has
shortened considerably, requiring less than one Gyr to go from $T_{\rm
eff}$ =
6500 K to 5500 K, and the SCZ deepens significantly (and rapidly, see
the nearly vertical tracks for $0.775 M_{\odot}$ and $0.725 M_{\odot}$  in
Figure 2).
Thus,
unless spallation production rates increase proportionately (and quite
significantly), $^6$Li that was created by surface spallation during the
(late)
main sequence will now dilute.  This is in {\it sharp} contrast to subgiant
evolution protostellar $^6$Li.  Whereas spallated $^6$Li will dilute as
the model
evolves {\it immediately} past the turnoff, since the only region of the
star containing $^6$Li is essentially the convection zone itself,
protostellar $^6$Li
will begin diluting only after the convection zone deepens past the entire
(protostellar) $^6$Li preservation region, which is orders of magnitude
larger
than  what $M_c$  is at turnoff (compare Figs. 4a and 4b).  This is the key
test
we
propose to discriminate between spallated and protostellar $^6$Li.

	Because of the potential significance of this test, we point out
evidence supporting the proposed depth of the Li preservation region.
The numerous observations of $^7$Li in halo subgiants of Pilachowski
{\it et al.}
(1993) follow closely the morphology of the standard DDK dilution
tracks
down to about 5000 K: the (subgiant) $^7$Li plateau, the onset of
subgiant
$^7$Li dilution, its progression, and its termination onto the diluted
plateau
(see Ryan and Deliyannis 1994).  This suggests that the {\it profile} of
the Li preservation region in stars may not be too different from that in
the
DDK models.  However, we must point out that we are not advocating
that the standard DDK models uniquely portray the true Li evolution in
halo stars, since other models with additional physics (e.g. rotational
mixing) might also have realistic $^7$Li profiles (and thus also be
consistent
with the data), but have different implications for protostellar Li.
(Indeed,
the decline of subgiant Li below 5000 K is inconsistent with standard
models, as is the decrease in
the $^{12}$C/$^{13}$C there, e.g. Deliyannis {\it et al.}
1994.)  The point is that these subgiant data corroborate the predicted Li
profile, and thus the depth to which protostellar $^7$Li is preserved,
which is a much deeper region than that contained in $M_c$ at turnoff.
Since both the temperature and
density at the base of the SCZ are still much too low to allow burning of
$^6$Li near that phase (and are not likely to be sufficiently high to do so
even
in view of the model uncertainties), it is likely that protostellar $^6$Li is
also
preserved to near its model depth, i.e. about twice as shallow as for
$^7$Li,
but still far deeper that the depth of SCZ near turnoff.  Therefore, if
$^6$Li is
observed in subgiants near the turnoff, protostellar $^6$Li should also
be
observed with similar abundance near the onset of subgiant $^6$Li
dilution at
5800 K, whereas surface spallated $^6$Li will have already diluted by
one or
more orders magnitude near 5800 K.

	Finally, we point out an interesting difference between dwarf and
subgiant $M_c$  isochrones: at a given $T_{\rm eff}$, the subgiant SCZ
is shallower
than the dwarf SCZ (e.g. Figure~3), so that for spallation rates independent
of
$T_{\rm eff}$, subgiants will have more spallated $^6$Li than dwarfs at
the same $T_{\rm eff}$.
(This is also what is expected for protostellar $^6$Li.)

\medskip
\noindent
{\it 2.3.  The Case of HD~84937}

	Literature estimates for the effective temperature of HD~84937 lie
near 6250 K, with a range from 6100 K to 6400 K.  If HD~84937 is a
dwarf, this corresponds to $M_c  = 1\times 10^{-3} M_{\odot}$, with a range
$3\times10^{-3}$ to
$3\times 10^{-4}M_{\odot}$ (using $\alpha = 1.5)$; if HD~84937 is a subgiant,
it
corresponds
to $ \sim 3\times 10^{-4}M_{\odot}$ with a range of $ \sim 7\times 10^{-5} -
1\times 10^{-3}M_{\odot}$, i.e. a factor of $3 - 4$
smaller.  These are our best ranges for the star, with uncertainty (from all
sources) perhaps of at least an order of magnitude.  To be
viable, surface spallation must be able to create an observable $^6$Li
abundance in at least this much mass.

\medskip
\centerline{3. NUCLEAR SPALLATION BY STELLAR FLARES}

Production of the light isotopes by stellar flares has been
previously investigated by several groups (eg. Ryter {\it et al.} 1970;
Canal {\it et al.} 1975;  Walker {\it et al}. 1985).
 The main aim of
these previous studies was to establish if such a mechanism could be a
 credible alternative to the
GCR hypothesis for the universal origin of the light isotopes.
The general consensus from these studies
was  that flare production (at least as the sole mechanism) was not a
viable
option.

However, in this work our aim is not as grandiose. Given the importance
of the  SLN  observation,
we simply wish to determine the
feasibility of flare production of $^6$Li  in HD~84937 -- and more
importantly
to determine observational tests which would allow such flare production
to
be discriminated against. For it is only if flare induced $^6$Li can be
safely
disregarded in such stars would it then be reasonable to draw any
cosmological implication from
the observational data.

Let us first consider the energetics of $^6$Li production in HD~84937.
For
the stellar flare
 we adopt
a source spectrum of the form
$$\Phi_i(E)=\alpha_i E^{-\gamma}  \ \ \ \  i={\rm p}, \ \alpha\>,
\eqno(3.1)$$
where $\alpha_i$ is a normalization constant, $E$ is the energy per nucleon,
 and  the spectral
index
$\gamma$ is taken to be in the regime $2-7$. Due to energy losses,
however,
the source spectrum can evolve as the particles are transported through
matter.
In terms of $\Pi$, the amount of matter traversed in atoms cm$^{-2}$,
 such
evolution of the spectra  can be approximated as
$$\Phi_i(E,\Pi)={\alpha_i E_i \over (E^2_i+2A_i\Pi)^{{\gamma +1\over
2}}}\>, \eqno(3.2)$$
where $A_i=2Z_i^2\times 10^{-21}$ MeV$^2 $atom$^{-1}$ cm$^{2}$
($Z$ being the
nuclear charge). Clearly, in the limit $\Pi<< E^2/2A$, the transported
spectra
collapses to the original source spectra. As discussed in Ryter {\it et al.}
(1970),
one can  make various approximations in order to estimate, $\eta$,
the energy required to form one atom.
  These various approximations
depend on the ionization level of the medium through which the flux
propagates, and
whether  the particles are re-accelerated after leaving the source.
For example, assuming no re-acceleration in a neutral plasma,
the energy required to form one light isotope
from a flux $\Phi_i(E)$ impinging on particles $j$
can be written as
$$\eta={\int^\infty_0\Phi_i(E) E dE \over
\int^\infty_0\biggl [\Phi_i(E) \sum_{j} {n_j\over
n_H}\int^E_{Q_j}{\sigma (E)\over \epsilon (E)}
dE' \biggr ]dE}\>, \eqno(3.3)$$
where $n$ is number density
($H$ refers to hydrogen),
 $\sigma(E)$ is the production cross section
(listed by Read and Viola 1984),
$Q_j$ the corresponding energy threshold, and  $\epsilon (E)=(1/n_H)
(dE/dx)$ is
the stopping power of the gas ($x$ being the length of the  particle path).
Tabulations of the stopping power in neutral matter can be found in
Barkas and Berger (1964),
whereas the stopping power in plasmas are discussed in detail by Spitzer
(1965).
For typical flare parameters, equations~(3.2) and (3.3) give $\eta\sim 1$ erg.

Of course, as already indicated, the actual values of $\eta$ depend on the
approximation made, and the
input flare parameters. The low energy cut-off in the flux spectra also
plays an important role.
With variation of these parameters, and allowance for
 the
presence of re-accelaration and non-zero ionization levels,
$\eta$ is typically found to lie in the range $1-1000$ ergs (Ryter {\it et
al.}
1970;
Canal {\it et al.} 1975; Canal {\it et al.} 1980). The lower values of
$\eta$
are usually associated with hard (lower $\gamma$) flares in hot
environments.

Adopting the stellar model parameters discussed in section~$2$, we can
 estimate the total energy, $E_T$, required for production of all the
$^6$Li
detected in HD~84937. For example,  assuming $\eta=1$ erg and
a convective envelope mass  $M_c\sim 10^{-3} M_\odot$, we find
$E_T\sim10^{42}$ ergs.
We can
put this value of $E_T$ in perspective. Assuming a
production timescale
of $10^9$ years,
the ``average''  required luminosity of a flare
 would be 1 part in $10^6$ -- in terms of HD~84937's luminosity.
Another way of putting this  is to consider the
flare activity required. Large solar flares typically provide $10^{32}$
ergs. Given
the evolutionary timescales, the number of such flares required   would
be $\sim 10$ per year for $E_T=10^{42}$~ergs.  However, it is  worth noting
that given the uncertainty in the
convective envelope mass, the required flare activity could be as low as
 one $10^{32}$ erg flare  per $100$ years.

Given that the required energy requirements
can at least in principle be  fulfilled, as
a working hypothesis
 let us assume flare induced
   proton and alpha-particle  interactions
with ambient alpha-particles and CNO nuclei in the
stellar atmosphere  are the main
source of  the $^6$Li observed  in HD~84937.
We now calculate the light isotopic ratios in HD~84937
based on this premise.

The number density $n_k$ ($k=^6$Li, $^7$Li, $^{9}$Be, $^{10}$B,
$^{11}$B)
 of each isotope produced by a flux of $i$ particles can
 be described by the differential equation
$${dn_k\over dt}= \sum_{j} \biggl [ \int^\infty_{Q_{j}} \Phi_i(E,\Pi)
\sigma(E)S_k(E)dE \biggr ] n_j(t)
\ \ \ . \eqno(3.4)
$$
The parameter  $S_k(E)$  represents the  survival probability of the
particle $k$ produced.
Adopting Fe and CNO abundances representative of HD~84937,
namely,
${\rm [Fe/H]}=-2.3$,
[O/Fe]$=0.5$,
[C/Fe]=[N/Fe]=$0$, and adopting $\alpha_p/\alpha_\alpha=10$, we solve
equation~(3.4) for a variety
of
flux spectra.  The calculated abundances are then convolved with the
stellar
models discussed in  section~2. This then gives predictions of the isotopic
abundances
in the stellar atmosphere as a function of $T_{eff}$. On the subgiant branch,
the calculated
number densities,
$n(^6{\rm Li})$ and $n(^7{\rm Li})$,
 are again normalized
by setting $n(^6{\rm Li})=n_o(^6{\rm Li})$ at
$T_{eff}=6250$K -- the observed values being
$n_o(^6{\rm Li})=6.6 \times 10^{-12}n_H$ and
$n_o(^7{\rm Li})=132 \times 10^{-12}n_H$ (SLN).
 The primordial $^7$Li value is then set at
  $n_p(^7{\rm Li })=n_o(^7{\rm Li })-20 R n(^7{\rm Li })$, $R$ being
the calculated
$^7$Li/$^6$Li ratio.
  It should be noted that this pre-spallation value may be different from
the
protostellar value due to stellar processing (see section 4), though in the
case of standard models near the turnoff the two would be almost
identical.

\medskip
\noindent
{\it  3.1.  Subgiants}

The two main parameters in our calculations are the spectral index
$\gamma$ and
the grammage $\Pi$.  For the purpose of calculation let us assume  the
values
$\gamma=3$ and $\Pi=0.1$ g cm$^{-2}$. We also assume that the flare
activity
remains constant.
Figure 4b shows the calculated $^7$Li/$^6$Li ratio as a function of
$T_{eff}$
for a star which  is in the subgiant phase of evolution.
 It can be
clearly seen that closer to the turn-off point (i.e. higher $T_{eff}$) the
$^7$Li/$^6$Li becomes significantly smaller, reaching values as low as
$\sim 5$.
Contrary to this, we can see that far from the turn-off point the
$^7$Li/$^6$Li ratio becomes larger, rapidly increasing beyond values of
$30$ for $T_{eff}$'s
cooler than about $6200K$.  The important point is, however, that as
discussed
in section 2.1, if the lithium isotopes have a protostellar origin
then the $^7$Li/$^6$Li ratio ($\sim 20$) should remain a constant in this
$T_{eff}$ region.

This difference in predictions, namely constant $^7$Li/$^6$Li toward
cooler
$T_{eff}$ (protostellar $^6$Li) versus increasing $^7$Li/$^6$Li toward
cooler $T_{eff}$
(spallated $^6$Li )  is large enough
to be observable by current techniques, and represents our key
discriminatory
test between flare and protostellar production.
We note that the test is robust
in the sense that different input parameters for the flux spectra (i.e.
different values
of $\gamma$ and $\Pi$) give very similar results to those shown in
Figure~4b.
The predicted trend (and size of the effect)
of the flare production scenario is independent of the input parameters.
This is essentially because, to first order, the controlling effect is the
size of the convective envelope at a given $T_{eff}$.

Variations in the flare activity with time and from star to star,
 can obviously affect the calculated yield. This would result
in a less uniform increase of the predicted $^7$Li/$^6$Li ratio as a
function of $T_{eff}$.   That is, there may be differences (a spread) in
$^6$Li abundances at a given $T_{eff}$.
Similarly, the likelihood of an age spread in the Galactic
halo (even at constant metallicity) suggests that stars could form with
different
initial (GCR produced) protostellar $^6$Li, again resulting in a spread of
$^6$Li at a given $T_{eff}$.
Nonetheless, the observational test would remain in an average sense  -- in
a sufficiently large sample.
Constancy (on average) of the
observed $^7$Li/$^6$Li ratio  with $T_{eff}$ on the subgiant branch
 would support a protostellar origin,
whereas increase of the same ratio (on average) at cooler $T_{eff}$
would support flare production.

\medskip
\noindent
{\it 3.2. Dwarfs}

A similar calculation could be carried out for the dwarf case.
However, here our ignorance of the flare parameters will play a  more
important role.
The reasons for this is that the predicted trend of the
$^7$Li/$^6$Li ratio is similar for both the flare and the
protostellar origins above
$T_{\rm eff}$ = 5800 ;  compare fig 4a and b.  This can be
seen from the discussions of
section 2.1, where it is discussed why protostellar  $^7$Li/$^6$Li ratio
will increase with decreasing $T_{eff}$ for dwarfs.  However, since the
mass of the convective envelope increases with  decreasing $T_{eff}$ for
dwarfs,
the
predicted $^7$Li/$^6$Li ratio from flare production
will likewise increase with decreasing $T_{eff}$.  As the trend of
the $^7$Li/$^6$Li ratio with $T_{eff}$ is then similar for both
production mechanisms,
any discriminatory test would rely solely upon the amount of flare
production
versus the amount of lithium destruction on the pre-main sequence.
Clearly, this would require detailed knowledge of the flare history
over the lifetime of the star -- knowledge of which eludes us.
There are also model uncertainties in evaluating the precise amount of
pre-main sequence $^6$Li destruction (DDK), further confounding
attempts to discriminate between scenarios using the dwarfs.

We see therefore, that unlike subgiants,
dwarfs are not a good testbed of the differing production scenarios.
Only if large $^7$Li/$^6$Li ratios are observed in dwarfs, and are
deemed
 incompatible with  the  calculated
pre-main sequence depletion factors, could indirect evidence for some
non
protostellar
origin be forthcoming.

For cooler dwarfs ($T_{\rm eff} < 5800$K), a $^6$Li
detection would point to a spallative origin since protostellar $^6$Li
would
have been completely destroyed.  However, though $^7$Li/$^6$Li is
much
smaller than the protostellar case, at  a value $> 100$ it may be too large
to allow
detection of spallative $^6$Li.

\medskip
\noindent
{\it 3.3.  B/Be ratios}

Although the  $^7$Li/$^6$Li ratios predicted above are largely
independent
of the flare parameters $\gamma$ and $\Pi$, this is  certainly not the
case for the predicted Beryllium and Boron yields. That is, the predicted
Li/Be and Li/B
ratios can vary widely, dependent on the input values of $\gamma$ and
$\Pi$. In
addition, these latter ratios are dependent on the assumed ratio of
alpha-particle to proton ratio in the flare spectra (since the lithium
isotopes
are mainly produced by $\alpha-\alpha$ fusion reactions at low
metallicity).
For practical purposes this translates into the fact that the flare mechanism
has little predictive power with regard to Li/Be and Li/B ratios. Although  the
recent
detection of
Be in HD~84937 by Boesgaard and King (1993) can be accounted for  by flare
production, in general
the Li/Be ratio predicted solely from flares can be very large. In such
circumstances,
additional production of Be and B from some other mechanism (eg GCR's) would be
necessary.
Although B has recently been detected in a few metal-poor halo stars
(Duncan {\it et al.} 1992); Edvardsson {\it et al.} 1994), as yet there has
been no detection of B in
HD~84937.

However, the tight predictive power of GCR spallation theory, does
allow
for
discriminatory tests to be made. This is better described in terms of the
B/Be ratio,
as this ratio is more independent of the alpha-particle to proton flux in the
flare,
and is free of the issues surrounding the primordial lithium value.
Our calculations show the B/Be ratio  predicted by a  flare mechanism
can readily reach values of $\sim 100$ for typical spectra.
Such an increase in the B/Be ratio with $\gamma$
is well known (eg. Walker {\it et al.} 1985), and can be understood from
considerations
of the energy dependence of the spallation cross section data. Simply
put,
spectra with
large $\gamma$'s give more weight to the low energy regime where the
lower threshold
of the B production reactions  lead to larger B/Be ratios. A more detailed
discussion of the energy dependence of B/Be is given by
Duncan {\it et al.} (1992).

The B/Be ratio predicted by GCR spallation theory, on the otherhand, is
constrained
to be  less than $20$ (somewhat higher values are possible but only at the
expense
of adopting contrived GCR spectra). This upper limit of $20$ arises from the
folding of the
cross-section
data and input initial abundances, with the present-day observed cosmic-ray
spectrum.
Although this measured spectrum could have evolved from an initially
different
spectrum prevalent in the early galaxy (Prantzos {\it et al.} 1993),
nuclear
destruction effects limit such evolution, and the predicted B/Be ratio  in
the
early
galaxy is essentially the same as that predicted using the present day
cosmic-ray spectrum
(Malaney and Butler 1993).
We  note, however, that Deliyannis and Pinsonneault (1990)
suggest that rotational stellar models may allow for slight Be (but no B)
depletion,
thereby increasing the B/Be ratio inferred from observation by $1.5$.
Also, we note the
work of Kiselman (1993), who cautions that non-LTE effects can  substantially
raise the value
of B/Be  inferred from observations of metal-poor stars. However, in their
non-LTE analysis of
HD~140283,
 Edvardsson {\it et al.} (1994) conclude that  the B abundance should be
increased by $\sim 0.5$ dex
relative to the LTE value, resulting in a B/Be range of $9-34$ and therefore
still
consistent with a GCR origin.
Only if it can be established from observation that   B/Be$> 20$
 ($>30$ if rotational stellar models are employed) in a halo star atmosphere,
 would there then be a strong case for some
production mechanism other than GCR spallation.

We caution here that neutrino-spallation processes may confuse the
issue.
Neutrino-spallation
 can  lead to production of B (Woosley
{\it et al}. 1990), and perhaps also to small yields of Be (Malaney 1992).
The predicted B/Be ratios from neutrino-spallation acting alone are
typically very large ($>200$).
Recently, Olive {\it et al.} (1994) have argued that GCR plus neutrino
spallation
leads to a relatively model independent prediction of B/Be$>50$ for
${\rm [Fe/H]}<-3$. In view of this,
 B/Be$>30$ at higher metallicities  (${\rm [Fe/H]}\sim -2$) would be the
key
indicator of some
flare-induced light isotope production.

\medskip
\centerline{4. DISCUSSION}

	So far we have presented a simplified picture of how flares might
be able to produce observable amounts of $^6$Li in turnoff halo stars,
and
how one might be able to distinguish between surface spallated $^6$Li
and
protostellar $^6$Li.  In this section we discuss how this picture might be
affected by various complications.

\medskip
\noindent
{\it 4.1.  Dependence of $M_c$  on Metallicity}

	It is hoped that $^6$Li observations will be attempted in a large
sample of halo dwarfs with a wide variety of metallicities.  It is therefore
relevant to examine how our results might depend on metallicity.  The
models of DDK show that both destruction of protostellar $^6$Li and
also
subgiant evolution of protostellar $^6$Li are only slightly dependent on
metallicity.  Fortunately, $M_c$  also depends only weakly on model
metallicity for halo metallicities, with the $M_c$  evolution pathway for
$Z=10^{-3}$ lying at most 0.2 dex above the $Z= 10^{-4}$ pathway at
their
widest point of separation, which is unimportant for our purposes here
(Figure~5).  The $M_c$  evolution pathway for $Z=10^{-5}$ nearly
coincides with
that for $Z=10^{-4}$.  Note that $Z=10^{-3}$ models have lower
turnoff $T_{\rm eff}$'s
than $Z=10^{-4}$ models of the same mass (turnoff age).  For
subgiants, the
dependence of $M_c$  on metallicity is more significant: at 6300 K,
$M_c$
differs by a factor of roughly $3$ between $Z=10^{-3}$ and $Z=10^{-
4}$, though the
difference effectively disappears at lower $T_{\rm eff}$.  On the other
hand,
fortuitously, the uncertainty in $M_c$  due to ambiguity in evolutionary
status
is smaller for model subgiants with $Z=10^{-3}$ than with $Z=10^{-
4}$.

\medskip
\noindent
{\it 4.2.  Dependence of $M_c$  on the Choice of Mixing
Length}

	Figure 6 shows that near the turnoff, $M_c$  depends strongly
on the
choice of mixing length: in going from $\alpha = 1.5$ to 1.1 at a
$T_{\rm eff}$ (at 17
Gyr), the SCZ becomes an order of magnitude shallower.
Unfortunately,
choice of appropriate value of $\alpha$ is fraught with uncertainty.  One
possible method is to take advantage of the sensitivity of $\alpha$ to
model
radius and the fact that the solar radius is known precisely.  This yields
$\alpha = 1.4$ for a solar model with the same treatment of physics as in
the
models of DDK (but there is nothing sacred about this value: models that
treat input physics or even convection differently may require a different
value for $\alpha$).  However, there is no guarantee that this value, even
if
appropriate for the sun, will be appropriate for halo dwarfs.  In fact,
reasons have been discussed as to why other values may be more
appropriate for halo dwarfs (Deliyannis and Demarque 1991; Chaboyer {\it et
al.} 1992).  It should be borne in mind also that mixing length (convection)
theory does have its limitations, and that the effective mixing length
could
well be a function of stellar mass, age, and other variables.  In short,
even
if we knew a star's $T_{\rm eff}$ perfectly, there is still great systematic
uncertainty
in (absolute) $M_c$  due to uncertainty in the treatment of convection.
However, even if convection can currently only be modelled
approximately, convection itself may be rather similar in metal poor stars
of similar $T_{\rm eff}$, so the models may be giving us a better
estimate of the
differences in $M_c$  between stars.

\medskip
\noindent
{\it 4.3.  Sensitivity to Model Physics}

	Deliyannis and Demarque (1991) caution that $M_c$  depends
sensitively on opacities and the treatment of other input physics, so a
comparison to other models might provide a useful indicator as to the
size
of the model uncertainties in $M_c$ .  For example, the models of
Proffitt and
Michaud (1991, PM) do treat some of the input physics a bit differently
(for discussion see Chaboyer  {\it et al.} 1992; Proffitt and
Vandenberg
1991).  Nevertheless, the standard dwarf $M_c$  isochrone that PM
show,
at 15 Gyr, ${\rm [Fe/H]} = -2.3$, and $\alpha = 1.5$, is nearly identical
to our
corresponding $M_c$  evolution pathway (for ${\rm [Fe/H]} = -2.3$,
$\alpha = 1.5$).
However encouraging, the comparison is not so straightforward; for
example, PM's value of $\alpha = 1.5$ is not solar calibrated (Proffitt
1994),
and there are some unavoidable differences in the input parameters as
well.  Nonetheless, it is not unreasonable to assume that the model
uncertainties in $M_c$  are closer to of order a factor of two or so, rather
than
 an order of magnitude; this accuracy suffices for the purposes of
this paper.  Support for this comes by comparing to the models of
Chaboyer (1994), which have made use of some improvements in the
input
physics (e.g. opacities).  The new $M_c$'s are indeed within a factor of
2-3.

\medskip
\noindent
{\it 4.4.  Additional physics: diffusion}

	The possibility that microscopic diffusion (e.g. gravitational
settling, thermal diffusion) has affected surface Li abundances in halo
stars has been discussed and debated (Michaud {\it et al.} 1984,
Deliyannis {\it et al.}
1990, Chaboyer {\it et al.} 1992).  Diffusion of $^6$Li should be quite
similar to
that for $^7$Li.  It has been argued that the $^7$Li abundances
observationally
constrain the effects of diffusion (acting alone) to be rather small
(Deliyannis and Demarque 1991).  Conceivably, diffusion can be more
important if its effects are masked by rotational mixing ($^7$Li then no
longer
tracks diffusion, since it can be mixed and destroyed); but in that case,
the
effects of rotational mixing are even more important (below).  In either
case, diffusion does not significantly alter our results.

\medskip
\noindent
{\it 4.5.  Additional Physics: rotational mixing}

	Recall that the predictions of standard stellar models are blatantly
contradicted by Pop I data (e.g. the Boesgaard Li gap and depletion of
Be
in F stars, the degree and timing of Li depletion in open clusters, the
dispersion of Li abundances observed in open clusters, the higher Li
abundances seen in short period tidally locked binaries).  In contrast,
rotational models do a much better job of explaining these (and other)
observations.  One must therefore consider the possibility that rotational
mixing has been important for Pop II stars as well, especially since
rotational models are able to explain more features of the Pop II
observations than do standard models.  The possibility that rotationally
induced mixing has affected the Li isotopes introduces complications.
The most obvious of such complications
would be a likely inconsistency with standard BBN; since if significant Li
depletion has taken place
in the halo stars,  the value of ${\rm Li}_p$ would be above that predicted by
standard BBN.
The recent quasar D observation (if taken at face value) also suggests an
inconsistency in standard BBN, but rotational mixing could conceivably
solve both problems through its effect on $^3$He on the giant branch, and the
resulting modifications to the D$+^3$He arguments (Deliyannis {\it et al.}
1994).

In the context of these rotational depletion models, one
explanation for the origin of any detected $^6$Li in halo stars would be
surface spallation -- rather than protostellar.
This would seem reasonable since one anticipates the  rotationally-induced
depletion factors
for $^6$Li to be larger than those for $^7$Li,  and if initially the
protostellar
$^6$Li was much less than protostellar $^7$Li, then very little $^6$Li would
still be
present in the stellar atmosphere.

 Of course, an alternative solution
could be that {\it both} the protostellar $^6$Li and $^7$Li were initially
higher
than currently observed, and that the  preferential destruction of $^6$Li
relative to $^7$Li results in the $^7$Li/$^6$Li ratio now observed in HD~84937.
The main point is that  rotational depletion (which
 is a very different process from
nuclear burning at a given temperature)   is more
analogous to a dilution process, where Li rich
material from above is mixed {\it slowly} with Li poor material from
below.  Thus, rotational depletion of $^6$Li need be only slightly larger
than
that for $^7$Li, depending on the rotational parameters and $M_c$  relative to
the sizes of each Li
preservation
region.  $^6$Li does depend more strongly
on
the rotational parameters than does $^7$Li.  So, for example, for an initial
angular momentum
$J_o$ just
slightly smaller than case J0 shown in DDK, $^7$Li depletion may be a
factor
of 7 - 8 and $^6$Li depletion only a factor of 25 - 40.  This would be
consistent with $A({\rm Li}_p$) $ \approx  2.9 $,
$^7{\rm Li (GCR)}  \approx ^6{\rm Li (GCR)}\approx  2.2$, giving current
observed
abundances of $^7$Li $\approx  2.1$ in
HD~84937 and $^6$Li a factor of 20 smaller, consistent with SLN.
Only slightly higher $J_o$ would affect $^7$Li very little but push
$^6$Li down to
unobservable levels, whereas only slightly lower $J_o$ could result in
even
lower $^7$Li/$^6$Li ratios.

  Further discussions of this point, and issues
as to whether  high $^6$Li (GCR) values are feasible within the context of GCR
models,
are beyond the scope of the present work. Suffice to say that
GCR models cannot readily reproduce the shallow slope of the
observed
Be-Fe relation (eg. see discussion in Malaney and Butler 1993),
and the introduction of more complex GCR models may open
up new
possibilities for interpreting $^6$Li detections in halo stars in the context
of
rotational depletion.
One could readily {\it contrive}
a new GCR model which predicts high   $^6$Li (GCR), while remaining consistent
with  all the available light isotope data at low metallicities.

Finally, we wish to point out that,
in the presence of rotationally-induced
depletion,
 our proposed test for distinguishing
between the different production scenarios is largely negated.
	Depletion of both protostellar and spallated $^6$Li would
continue to
the present, though would be at its weakest in the last several Gyr.  Since
protostellar $^6$Li depletion is larger than that of $^7$Li, it is possible
that $^6$Li
would be preferentially depleted on the subgiant branch, perhaps
increasing $^7$Li/$^6$Li by a factor of two or so there.  This lessens
the
difference between the subgiant spallated $^6$Li vs. protostellar $^6$Li
cases
discussed above.  Furthermore, since protostellar $^6$Li depletion is
more
sensitive to rotational parameters, such as $J_o$,
than is $^7$Li, rotational models would predict a large variety
of $^7$Li/$^6$Li
in subgiants, with $^6$Li possibly detectable in some but not in others.
Unfortunately, these factors make it more difficult to distinguish between
alternative scenarios.

	Detections of $^6$Li in halo dwarfs (assuming they could be
demonstrated to be protostellar and NOT spallative in origin) could,
nonetheless, offer some useful constraints for rotational models.  The
level in HD~84937, for example, argues against models with higher
initial
angular momentum (and wind law and other parameters as assumed),
such that $^7$Li/$^6{\rm Li} > 100$.  The higher the detected $^6$Li
abundance, the
stronger the constraints, at least in those stars.

\medskip
\centerline {5. CONCLUSIONS}

We have investigated in detail  the possibility that flare activity on the
surface
of halo stars may lead to observable quantities of $^6$Li -- thereby impacting
on the conclusions to be drawn from any observation of this isotope in
a metal-poor atmosphere. Our calculations have been largely on the premise that
standard BBN is
an accurate description of primordial  isotope production, and that standard
stellar evolution models are applicable. We found that  energetic flares
could account for the quantity of $^6$Li observed in HD~84937,
 therefore limiting the reliability of cosmological implications drawn from
 this type of observation.
As such, we proposed a future observational test which would
allow a discrimination against flare production to be made.

A critical quantity in our calculations was the  mass of the surface convective
zone, $M_c$.
We found that on an isochrone, $M_c$  decreases steeply with
increasing $T_{\rm eff}$,  decreases steeply with age, and was only weakly
dependent on metallicity.
  $M_c$  deepens rapidly on the subgiant
branch,
but for lower metallicities remains a bit shallower than $M_c$  in dwarfs
of the
same $T_{\rm eff}$.  Assuming surface spallation production rates are
approximately independent of time and $T_{\rm eff}$, flare-spallated
$^6$Li is most
likely to be observed in the hottest, oldest dwarfs, and preferentially in
subgiants.  More evolved subgiants will have diluted their
flare-spallated $^6$Li, but will have preserved their protostellar $^6$Li.
Given a statistically meaningful sample of observations of the $^7$Li/$^6$Li
ratio
in metal-poor halo stars,  contamination by flare production could be
discriminated against, and important cosmological implications could  be more
safely inferred.  If rotational Li depletion is applicable, then
complications are introduced both in applying our test to distinguish
between flare-spallated $^6$Li  and protostellar $^6$Li, and in general in
interpreting the $^6$Li abundances.

\medskip
\noindent
{\it Acknowledgements.}

\medskip
\noindent
We thank B. Chaboyer for useful discussions.
C.P.D.  gratefully acknowledges support for this work by NASA
through
grant HF P 1042.01 P 93A awarded by the Space Telescope Science
Institute which is operated by the Association of Universities for
Research
in Astronomy, Inc., for NASA under contract NAS5-26555.  This
research has made use of the Simbad database, operated at CDS,
Strasbourg, France.

\vfill
\eject
\centerline{REFERENCES}

\item{}Barkas, W. H., \& Berger, M. J., 1964, NAS-NRC Pub. 113,
Report , 39,
103.

\item{}Boesgaard, A. M., \& King, J. R. 1993, AJ, 106, 2309

\item{}Canal, R., Isern, J., \& Sanahuja, B., 1975, ApJ, 200, 646.

\item{}Canal, R., Isern, J., \& Sanahuja, B., 1980, ApJ, 235, 504.

\item{}Carswell, R. F., {\it et al.}, 1994, MNRAS, in press.

\item{}Chaboyer, B. 1994, submitted to ApJL.

\item{}Chaboyer, B. \& Demarque, P. 1994, submitted to ApJ.

\item{}Chaboyer, B., Deliyannis, C. P., Demarque, P., Pinsonneault,
M. H., \& Sarajedini, A. 1992,  ApJ, 388,  372.

\item{}Demarque, P., Deliyannis, C. P., \& Sarajedini, A., 1991,
Ages of
Globular Clusters, in NATO Advanced Research Workshop
"Observational Tests of Inflation, Durham, England, eds. T. Shanks et
al., (Kluwer, 1991), p. 111.

\item{}Deliyannis, C. P. 1990, Ph.D. Thesis, Yale University.

\item{}Deliyannis, C. \& Pinsonneault, M., 1990,  ApJL.,  365, L67.

\item{}Deliyannis, C. P., \& Demarque, P. 1991, ApJL, 370, L89.

\item{}Deliyannis, C. P., Demarque, P., \& Kawaler, S. 1990, ApJS,
73, 21.

\item{}Deliyannis, C. P., Demarque, P., Kawaler, S. D., Krauss, L.
M., \&
Romanelli, P. 1989, Phys. Rev. Letters, 62, 1583.

\item{}Deliyannis, C. P., Demarque, P., \& Pinsonneault, M. H.
1994, in prep.

\item{}Deliyannis, C. P., Pinsonneault, M., \& Duncan, D. 1993, ApJ,
414, 740.

\item{}Dimopoulos, S., Esmailzadeh, R., Hall, L. J., \& Starkman, G. D.
1988, ApJ, 330, 545.

\item{}Duncan, D. K.,  Lambert, D. L., \& Lemke, M., 1992, ApJ, 401, 584.

\item{}Edvardsson, B., Gustafsson, B., Johansson, S. G., Kiselman, D., Lambert,
D. L.,
Nissen, P.E., \& Gilmore, G., 1994, Astron. Astrophys. in press.

\item{}Feltzing, S., \& Gustafsson, B., 1994, ApJ, 423, 68.

\item{}Fields, B., Olive, K. A.,  \& Schramm, D. N., 1994, Fermilab preprint
94/010-A.

\item{}Gilmore, G., Gustafsson, B., Edvardsson, B., \& Nissen, P.E., 1992,
Nature, 357 379.

\item{}Kiselman, D., 1994, Astron. Astrophys. in press.

\item{}Krauss, L. M., \& Romanelli, P., 1990,  ApJ,  358, 47.

\item{}Lee, Y-W., Demarque, P., \& Zinn, R., 1990, ApJ, 350, 155.

\item{}Malaney, R. A. 1992,  ApJL,  398,  L45.

\item{}Malaney, R. A., \& Butler, M. N., 1993, ApJL, 407, L73.

\item{}Malaney,  R. A., \& Mathews, G. J., 1993, Physics Reports, 229, 145.

\item{}Michaud, G., Fontaine, G., \& Beaudet, G., 1984, ApJ, 282,
206.

\item{}Olive, K. A., Prantzos, N., Scully, S., Vangioni-Flam, E., 1994, ApJ, in
press.

\item{}Pilachowski, C. A., Sneden, C., \& Booth, J. 1993, ApJ, 407,
699.

\item{}Pinsonneault, M. H., Deliyannis, C. P., \& Demarque, P. 1992, ApJS,
78, 179.

\item{}Prantzos, N., Casse, M. \& Vangioni-Flam, E.,  1993,  ApJ, 403, 630.

\item{}Proffitt, C. R. 1994, private communication.

\item{}Proffitt, C. R., \& Vandenberg, D. A., 1991, ApJS, 77, 473.

\item{}Proffitt, C. R.,  \& Michaud, 1991, ApJ, 371, 584 (PM).

\item{}Read, S. \& Viola, V., 1984,  Atomic \& Nuc. Data
Tables, 31, 359.

\item{}Ryan, S. G. \& Deliyannis, C. P. 1994, submitted to ApJ.

\item{}Ryter C., Reeves, H., Gradsztajn, E., \& Audouze, J.,  1970, Astron.
Astrophys. 8, 389.

\item{}Sarajedini, A. \& Demarque, P. 1990, ApJ, 365, 219.

\item{}Searle, L., \& Zinn, R. 1978, ApJ, 225, 357.

\item{}Smith, V. V., Lambert., D. L., \& Nissen, P. E., 1993, ApJ., 408, 262
(SLN).

\item{}Smith, M. S., Kawano, L. H., \& Malaney, R. A., 1993, ApJS, 85, 219.

\item{}Songaila, A., Cowie, L.L., Hogan, C.J., \& Rugers, M. 1994, Nature,
368, 599.

\item{}Spitzer, Jr., L., 1965,  Physics of the Fully Ionized Gases,
Interscience-New York.

\item{}Steigman, G., \& Walker, T. P.,  1992, ApJL, 385, L13.

\item{}Straniero \& Chieffi,  1991,    ApJS, 76, 525.

\item{}Thorburn, J. A. 1994, ApJ, 421, 318.

\item{}Vandenberg, D. A., Bolte, M. \& Stetson, P. B. 1991,
AJ, 100, 445.

\item{}Walker, T. P., Mathews , G. J. \& Viola, V. E. , 1985,
ApJS, 76, 525.

\item{}Walker, T. P., Steigman, G., Schramm, D. N., Olive, K. A., \&
        Kang, H.-S., 1991,  ApJ,  376, 51.

\item{}Woosley, S. E., Hartman, D. H., Hoffman, R. P., \& Haxton, W. C., 1990,
ApJ, 356, 272.

\vfil
\eject

\centerline {FIGURE CAPTIONS}

\medskip
\noindent
Figure 1: Color magnitude diagram of M92, with isochrones from
Demarque, Deliyannis, and Sarajedini (1991) and data from Stetson and
Harris (1988).  Two possible positions for the evolutionary status of HD
84937 are shown, based on a B-V color of 0.40: dwarf (A) and subgiant
(B).

\medskip
\noindent
Figure 2: Evolution of the mass $M_c$  contained in the surface
convection
zone (SCZ) for ${\rm [Fe/H]}= -2.3$ ($Z=10^{-4}$) and $\alpha =
1.5$, for masses
from 0.650 to $0.800 M_{\odot}$  in steps of 0.025 (some masses
labeled), as a
function of age.  X's mark the ZAMS.  For clarity, pre-main sequence
and
subgiant evolution is shown for only a few model sequences.

\medskip
\noindent
Figure 3: Evolution of $M_c$  in the $T_{\rm eff}$ - $M_c$  plane for
$Z=10^{-4}$ model
sequences.  Arrows indicate the direction of evolution.  For clarity, pre-
main sequence and post-turnoff evolution is shown for only a few model
sequences.  The ZAMS (shown in diamonds for each model stellar mass)
also defines the main sequence $M_c$  evolution pathway.  After the
ZAMS,
concave down curves show main sequence evolution while concave up
curves show subgiant evolution.  The numbers indicate stellar mass in
solar masses.

\medskip
\noindent
Figure 4a: Predicted evolution of protostellar $^7Li$/$^6$Li, at 17 Gyr (see
text
for parameters and normalization), showing the results of
pre-main sequence Li destruction.

\medskip
\noindent
Figure 4b :  Predicted trend of the  $^7Li$/$^6$Li ratio from flare activity as
a function of
$T_{\rm eff}$ for both the subgiant and the dwarf phases of evolution.

\medskip
\noindent
Figure 5: Comparison of dwarf isochrones (concave down) and subgiant
evolutionary tracks (concave up) of different metallicities ($Z= 10^{-
4}$, $18$ Gyr; $Z= 10^{-3}$, $17$ Gyr).

\medskip
\noindent
Figure 6: 17 Gyr dwarf isochrones for two different choices of mixing
length, at $Z=10^{-4}$.
\bye